\documentclass[11pt]{article} 
\usepackage{hyperref} 
\pdfoutput=1 
 
\title{Quantitative Flow Field Imaging about a Hydrophobic Sphere Impacting on a Free Surface} 
\author{Roderick R. La Foy, Tadd T. Truscott, \&\\   Alexandra H. Techet \\
\vspace{2pt} \\
The Department of Mechanical Engineering,\\
Massachusetts Institute of Technology
Cambridge, MA 02139, USA}
\begin{document}

\maketitle 
 
%% The abstract (in this file, and that submitted as text to arXiv) should include the exact phrase 
%% "fluid dynamics video" or "fluid dynamics videos" 
 
\begin{abstract} 
This fluid dynamics video shows the impact of a hydrophobic sphere impacting
a water surface. The sphere has a mass ratio of $m^* =\rho_s/\rho_w = 1.15$, a
wetting angle of $\alpha$~=~110$^\circ$, a diameter of 9.5~mm, and impacts the
surface with a Froude number of $Fr = U_o/\sqrt{gd} = 9.2$. The first sequence
shows an impact of a sphere on the free surface illustrating the formation of
the splash crown and air cavity. The cavity grows both in the axial and radial
direction until it eventually collapses at a point roughly half of the distance
from the free surface to the sphere, which is known as the pinch-off point. The
second set of videos shows a sphere impacting the free surface under the same
conditions using Particle Image Velocimetry (PIV) to quantify the flow field. A
laser sheet illuminates the mid-plane of the sphere, and the fluid is seeded
with particles whose motion is captured by a high-speed video camera. Velocity
fields are then calculated from the images. The video sequences from left to
right depict the radial velocity, the axial velocity, and the vorticity
respectively in the flow field. The color bar on the far left indicates the
magnitude of the velocity and vorticity. All videos were taken at 2610~fps and
the PIV data was processed using a 16~$\times$~16 window with a 50\% overlap.
\end{abstract}

% main text 
 
\section{Information} 
 
%The {\em hyperref} package is used to make links to the videos. 
%%  The format is:  \href{URL of video}{name that will appear in the text} 
 
Two sample videos are 
\href{http://ecommons.library.cornell.edu/bitstream/1813/11485/3/LaFoy_Truscott_GFM2008_mpg1.mpg}{(Video 1 mpeg-1)} and \href{http://ecommons.library.cornell.edu/bitstream/1813/11485/2/LaFoy_Truscott_GFM2008_mpg2.m2v}{(Video 2 mpeg-2)}.
 
\section{References}
\noindent
Truscott, T.T. \& Techet, A.H. ``Water-entry of Spinning Spheres''{\it J Fluid Mech.} Under Review 2008.\\
\\
\noindent
Truscott, T.T. \& Techet, A.H. ``Cavity formation in the wake of a spinning sphere impacting the free surface," {\it Physics of Fluids}, v. 18, no. 9, p. 18, September 2006.

\end{document}